\magnification 1200
\centerline{\bf Nucleosynthesis  in a simmering universe }
\vskip 3cm
\centerline {Daksh Lohiya, Annu Batra}
\centerline {Shobhit Mahajan, Amitabha Mukherjee}
\centerline {Department of Physics \& Astrophysics, University of Delhi,}
\centerline {Delhi 110 007, India}
\centerline {email: dlohiya@ducos.ernet.in}
\vskip 3cm
\centerline {\bf Abstract}
\vskip 1cm

     Primordial nucleosynthesis is considered a success story of the 
standard big bang (SBB) cosmology. The cosmological and elementary particle
physics parameters are believed to be severely constrained by the 
requirement of correct abundances of light elements. We explore
nucleosynthesis  in a class of models very different from SBB. In these
models the cosmological
scale factor increases linearly with time right through the period during 
which nucleosynthesis occurs till the present. 
It turns out that weak interactions remain in
thermal equilibrium upto temperatures which are two orders of 
magnitude lower than the corresponding (weak interaction decoupling)  
temperature in SBB. Inverse
beta decay of the proton can ensure adequate production of several light 
elements while producing primordial metalicity much higher than that 
produced in SBB. Other attractive features of these models are the 
absence of the horizon, flatness and the age problems and consistency
with classical cosmological tests. 

\vfil\eject

	Early universe nucleosynthesis is regarded as a major
``success story'' of the standard big bang (SBB) model. The 
results look rather good and the 
observed light element abundances are used to severely
constrain cosmological and particle physics parameters. 

	However, there is
no object in the universe that has quite the abundance [metalicity]
of heavier elements as is produced in the 
``first three minutes'' (or so)
in SBB. One relies heavily on success of some kind of
re - processing, much later in the history of SBB, 
to get the low observed  
metalicity in [eg.] old clusters and inter - stellar clouds. This could
[for instance] be in the form of a generation 
of very short - lived type III stars. Large
scale production and recycling of metals through such exploding early
generation stars leads to verifiable observational 
constraints. Such stars would be visible as 27 - 29 magnitude stars
appearing any time in every square arc - minute of the sky.
Serious doubts have been expressed on the existence and
detection of such signals [1].

	 Of late [2],   
observations have suggested the need for a careful scrutiny and a 
possible revision of the status of SBB nucleosynthesis from reported
high abundance of $^2D$ in several $Ly_\alpha$ systems. Though the 
status of these observations is still a matter of debate, and  
[assuming their confirmation-], attempts to
reconcile the cosmological abundance of deuterium and the number of 
neutrino generations within the framework of SBB are still on, we feel that
alternative scenarios should be explored. 

	Surprisingly,  a class of models 
radically different  from the standard one has a promise of
producing the correct amount 
of helium as well as the  metalicity observed in low metalicity objects. 
This paper is a status report on our ongoing efforts to study the 
cosmological implications of a class of models in which the cosmological scale
factor $R(t)$ varies linearly with time. The basic argument is quite
straightforward and goes along the lines of STD nucleosynthesis, 
summarised as follows: 

	A crucial assumption in the standard model is the existence of 
thermal equilibrium at temperatures around $10^{12}K$ or $100 MeV$.
At these temperatures, the universe is assumed to consist of leptons, photons 
and a contamination of nucleons in thermal equilibrium. The ratio of weak 
reaction rates of leptons to the rate of expansion of the universe 
(the Hubble parameter) below $10^{11}K$ (age $\approx .01 $ secs) 
goes as (see eg.[3]):
$$ 
r_w \equiv {\sigma n_l\over H} \approx ({T\over 10^{10}K})^3 \eqno{(1)}
$$
At these temperatures, the small nucleonic  
contamination begins to shift towards more protons and fewer
neutrons because of the n-p mass difference. 
By $10^{10}K$ i.e. $T_9 \equiv 10$, $r_w$ falls below unity, consequently, 
the weak interactions fall out of equilibrium and the
the neutrinos  decouple. The distribution function 
of the $\nu$'s however
maintains a Planckian profile as the universe expands. At 
$ 5\times 10^{9}K$ (age of about  4 secs), $e^+,e^-$ pairs
annihilate. The neutrinos having decoupled, all the entropy of
the $e^+, e^-$ before annihilation, goes to heat up the photons -
giving the photons some 40\% higher temperature than the temperature
corresponding to the neutrino Planckian profile. The decoupling
of the neutrinos and the annihilation of the $e^+, e^-$ ensures the
rapid fall of the neutron production rate 
$\lambda(p\longrightarrow n)$ in comparison to the 
expansion rate of the universe. n/p ratio freezes to about 
1/5 at this epoch. 
This ratio now falls slowly on account of decay of
free neutrons. Meanwhile nuclear reactions and photo -
disintegration of light nuclei ensure a dynamic buffer of light elements
with abundances roughly determined by nuclear statistical 
equilibrium (NSE). Depending on the baryon-entropy ratio, at a critical 
temperature around $T_9 = 1$, deuterium concentration is large enough 
for efficient evolution of a whole network of reactions leading up to the 
formation of the most stable light nucleus, viz. $^4He$. This
is the characteristic temperature at which $^2D$ conversion
into other nuclei becomes 
a more efficient channel for the destruction of neutrons than 
neutron decay. At slightly lower 
temperatures, deuterium depletion rate becomes small compared 
to the expansion rate [4] resulting in residual 
abundances of deuterium and $^3He$.  Elaborate numerical codes have 
been developed [5] to describe the evolution
of this phase. The abundances of deuterium, helium - 3,
helium - 4 and lithium - 7 can be used to constrain the baryon -
entropy ratio, the number of light particles around and the neutrino
chemical potential. The primordial metalicity obtained is rather low
and
one does not see any astrophysical object with metalicity (abundance
of lithium - 8 and heavier elements) as low as that predicted by
primordial synthesis alone. The oldest objects are believed to be
globular clusters. The metalicity reported in these systems is much
higher than accounted for by SBB and much too low in comparison with
that found in the atmosphere of population I stars and interstellar
gas. Special 
reprocessing and metal enrichment is suggested at a redshift of
10 to 5. No unambiguous experimental signal to this effect has been  
reported so far [1]. Consistency of the light element abundances 
in SBB, moreover, is ensured only if the baryonic matter density is some 
two orders of magnitude less than the closure density. This is regarded 
as a respite in SBB. Using the rest of the (non-baryonic) matter in a 
suitable combination of hot and cold dark matter (with possibly a 
small cosmological constant also thrown in) to build up 
large scale structures in cosmology has developed into an industry. 
The current status is not completely satisfactory.
In particular, the age estimates of globular clusters are uncomfortably high
in comparison with the age of the universe as set by 
conservative estimates.

	Motivated by the above, we explore the possibility of obtaining a 
consistent scenario for nucleosynthesis in a class of models which are
radically different from the standard one. In particular, we consider a 
cosmological model in which, right through the epoch when 
$T \approx 10^{12}K$ and thereafter, the scale factor $R(t)$ increases 
as $t$ (- the age of the universe). 
The linear evolution of the scale factor ensures a horizon-free 
cosmology. 
We shall later describe models in which such a scaling is possible. 
With such linear scaling, the present value of the scale parameter, i.e.
the present epoch  $t_o$, is exactly determined by the present
Hubble constant $H_o = 1/t_o$. The scale factor and the temperature of
radiation are related by  $RT \approx$ constant with effect from temperatures
$\approx 10^9K$. This follows from the stress energy conservation
and the fact that the baryon - entropy ratio does not
change after $kT \approx m_e$ (the rest mass of the electron). 
From present age and effective CMB temperature ($2.7K$). one finds 
the age of the 
universe when $T \approx 10^{10}K$ to be of the order of a few years. The
universe takes some $10^3$ years to cool from $10^{10}K$ 
to $10^8K$. The rate of expansion of the universe is about $10^7$ times 
slower than the corresponding rates for the same temperature in standard 
cosmology. This makes a crucial [big] difference and in fact
implies that the standard story does not go through. 

      The process of the neutrinos falling out of thermal equilibrium, for 
example, is determined by the rate of $\nu$ production per charged lepton:
$$ 
\sigma_{wk} n_l/c^6 \approx g_{wk}\hbar^{-7}(kT)^5/c^6\eqno{(2)}
$$
and the expansion rate of the universe [$H = 1/t$]. Here
$g_{wk} \approx 1.4\times 10^{-45}$ erg- $cm^3$. For 
$kT > m_\mu,~ T > 10^{12}K$
$$ 
\sigma_{wk} n_l/H \approx [{T\over {1.62\times 10^{8}K}}]^4 \eqno{(3)}
$$
Here we have normalised the value of $RT = tT =$ constant
from the value $H_o$ = 55 km/sec/Mpc for the Hubble constant -
corresponding to $t_o \approx 18\times 10^9$ years. 
[$tT_9 \approx 2.5\times 10^9$]. Increasing 
$H_o$ by a factor of 2 would merely lead to a change in the 
denominator on the right side of eqn.[3] to  $1.8\times 10^{8}K$.
When $kT < m_\mu$, the number density of muons is reduced by 
a factor $[exp(-m_\mu/kT)]$. Consequently, the rates of 
weak interactions involving muons get suppressed to
$$ 
\sigma_{wk} n_l/H \approx [{T\over {1.62\times 10^{8}K}}]^4 
exp[-{10^{12}K\over T}] \eqno{(4)}
$$

The corresponding rates in the standard model are:
$$ 
\sigma_{wk} n_l/H \approx [{T\over {10^{10}K}}]^3 \eqno{(5)}
$$
for $kT > m_\mu$, and
$$ 
\sigma_{wk} n_l/H \approx [{T\over {10^{10}K}}]^3 
exp[-{10^{12}K\over T}] \eqno{(6)}
$$
for $kT < m_\mu$. This would lead to the weak interactions
maintaining the $\nu$'s in thermal equilibrium to temperatures down 
to $1.62\times 10^{8}K$. The entropy
released from the $e^+ e^-$ annihilation heats up 
all the particles in equilibrium. Both the neutrinos and
the photons would therefore get heated up to the same temperature.
The temperature then scales by $RT =$ constant as the 
universe expands. The relic neutrinos and the photons 
(the CMBR) would therefore have the same Planckian profile
($T \approx 2.7K$) at present. (The photon number does not
significantly change at recombination for a low enough baryon
- entropy ratio). This is in marked contrast
to the standard result wherein the neutrino temperature
is predicted to be lower than the photon temperature.
The nuclear reaction rates are simply given by the 
expressions:
$$ 
\lambda(n\longrightarrow p) = A\int(1 - {m_e^2\over 
(Q+q)^2})^{1/2}(Q+q)^2q^2dq
$$
$$
\times (1 + e^{q/kT})^{-1}(1 + e^{-(Q+q)/kT})^{-1}\eqno{(7)}
$$
$$ 
\lambda(p\longrightarrow n) = A\int(1 - {m_e^2\over 
(Q+q)^2})^{1/2}(Q+q)^2q^2dq
$$ 
These rates have the ratio determined by the neutron - proton 
mass difference $\equiv Q \approx 15$ [in units $k = T_9 = 1$]:
$$
{\lambda(p\longrightarrow n)\over 
{\lambda(n\longrightarrow p}} = exp(-{Q\over T_9})\eqno{(8)}
$$
The rate of expansion of the universe at a given 
temperature being much smaller than that in the standard scenario, 
the nucleons are expected to be in thermal equilibrium with the ratio $X_n$ 
of neutron number to the total number of all nucleons
given by:
$$ 
X_n = {\lambda(p\longrightarrow n)\over 
{\lambda(p\longrightarrow n) + \lambda(n\longrightarrow p)}}
= [1 + e^{Q/T_9}]^{-1}\eqno{(9)}
$$
   
     As in the standard model, 
Deuterium burning into light elements becomes the 
more efficient channel for neutron destruction than neutron decay
at a temperature $T_9 \approx 1$ and nucleosynthesis commences. 
[This result follows from a numerical integration of the 
Boltzmann - rate - equations and was done by using Wagoner's [6] 
prescription]. 
At this temperature, one sees from eqn(9) that   
there are hardly any neutrons left.
However weak interactions
have not frozen off and inverse beta decay can convert protons 
into neutrons till temperatures down to $\approx 10^8K$. 
The baryonic content of the universe at 
$T_9 \approx 1$ is constituted by protons (mainly), some neutrons (less 
than 1\% ) and a buffer of light elements in NSE. 
The strength of the
buffer is enhanced by fresh neutron formation by the inverse beta
decay of the proton and its capture into the buffer by the pn
reaction. The buffer depletes by either: (i) the photodisintegration 
of any light element constituting the buffer followed by the decay 
of the resulting neutron before it can be recaptured into the 
buffer by the pn reaction; or (ii) the formation of $^4He$ which
is the most stable nucleus at these temperatures. Once helium 
formation becomes more efficient than neutron decay, most subsequent 
neutrons formed would precipitate into $^4He$.
This critical epoch of commencement of $^4He$ precipitation
is sensitive to the baryon-entropy
ratio. If the ratio of number of protons 
that convert into neutrons after this epoch, to the total baryon number of
the universe is roughly 1/8, we would get the observed $\approx 25\% ~^4He$.
To see this in a little more detail: eqn[8] implies:
$$
{{\lambda(p\longrightarrow n)}\over {\lambda(n\longrightarrow p)}}
= exp(-{Q\over kT})\approx e^{-15/T_9}\eqno{(10)}
$$
If $\tau$ is the neutron life time, eqn(10) gives:
$$
{\dot Y_p} \approx - {1\over \tau}e^{-15/T_9}Y_p \eqno{(11)}
$$
This is exactly integrated, starting from a temperature $T_{9o}$, 
to give:
$$
Y_p \approx Y_{po}exp[-{10^9\over 15\tau}e^{-15/T_{9o}}]
$$
$Y_{po} - Y_p$ is the number of protons converted to neutrons. 
If all these protons are converted into neutrons [i.e. $T_{9o}$
is the temperature at the epoch of $^4He$ precipitation
as described above],
the amount of helium is just:
$$
Y_{He} \approx 2[1 - exp[-{10^9\over 15\tau}e^{-15/T_{9o}}]]\eqno{(12)}
$$
This is  $\approx 24\%$  for $T_{9o} \approx 0.9$.
This simply translates 
into an appropriate requirement on the baryon-entropy ratio. 
Fortunately one has an extremely user friendly code [5]
that we modified 
to suit the taxing requirements of the much stiffer rate equations
that we encounter in our slowly evolving universe. To get 
convergence of the rate equations  for 26 nuclides and a network of
88 reactions [as given in Kawano's code], we executed some 500 iterations 
at each time step. An additional (89th) reaction
(the pp reaction): 
$$ 
p + p \longrightarrow D + e^+ + \nu \eqno{(13)}
$$ 
does not decouple on account of the slow expansion of the universe
and was incorporated in the code. 
The results for different values of $\eta$ are described in
table I.  We find consistency with the $^4He$ abundances for
$\eta \approx 10^{-8} $. The metalicity produced is 8 orders of 
magnitude greater than the corresponding value one gets in the
early universe in the Standard model. This is
also a consequence of the slow expansion in this model. A 
locally higher $\eta$ in an inhomogeneous model can further enhance
metalicity.

    To get the observed abundances of light elements besides $^4He$, 
one would have to fall back upon a host of other mechanisms 
that were being explored in the SBB in the pre - 1976 days.
The most popular processes are: (i) nucleosynthesis by secondary 
explosions of super massive objects [6], (ii) nucleosynthesis in 
inhomogeneous models, (iii) effect of inhomogeneous n/p ratios
as the universe comes out of the QGP phase transition, (iv)
spallation of light nuclei at a much later epoch.
It is easy to rule out the survival of $^2D$ by the processes (ii) 
and (iii) while the process (i) requires very special initial 
conditions. It also shares a common difficulty with process (iv),
viz.: the production of $^2D$ to the required levels is
possible but it is accompanied by an overproduction of lithium.
Any later destruction of lithium in turn completely destroys
$^2D$. Within the framework of the cosmological evolution that
we are exploring here, we find the best promise in a model that
would combine (ii) and (iv). Table 1 displays the extreme
sensitivity of $^4He$ production to $\eta$. In an inhomogeneous model
with a spatially varying $\eta$, there would hardly be any 
$^4He$ production in a region with $\eta$ lower by (say) a  
factor of two. Thus we can have proton rich clouds 
in low density regions and $^4He$ and metal rich clouds in
the higher density regions. The spallation of the former
on the later, at a subsequent [cooler] epoch, would produce 
$^2D$ without the excess production of lithium [7] as lithium forms 
primarily from spalling $^4He$ over $^4He$.

	We feel that one should be able to dynamically account
for such conditions within the framework of models we 
outline in the conclusion.

	With $R = t$, the expansion rate 
does not depend on the background 
density and thus nucleosynthesis is independent of the number of neutrino 
species or for that matter to any other (particles) 
extra degrees of freedom. 
The age of this universe (defined as the time elapsed from the hot
epoch to the present) would be exactly $50\%$ higher than the
SBB age determination, $2/3H_o$, from the Hubble
parameter. 

\centerline{\bf Conclusion}

      The purpose of the article is to show that a class of
cosmological models can not to be discarded away on account of
SBB nucleosynthesis constraints.
In any model in which the rate of expansion of the universe is low
enough to keep weak interactions in equilibrium at temperatures 
lower than the $^4He$ precipitating temperature, inverse beta decay
can lead to adequate $^4He$ and metal production. Further, in
principal, it is possible to produce $^2D$ by spallation of hydrogen
rich clouds over a $^4He$ - metal rich medium at a later epoch.

	 We finally address the issue of realising the linear evolution 
within the framework of a Friedman cosmology. Such an evolution
can be accounted for in a universe dominated by `K - matter' [8] for which 
the density scales as $R^{-2}$. 
The Hubble diagram (luminosity distance-redshift 
relation), the angular diameter distance - redshift relation and the galaxy 
number count-redshift relations do not rule out such  
a ``coasting'' cosmology [8,9].
However, if one requires
this matter to dominate even during the nucleosynthesis era, the K -
matter would almost close the universe. There would hardly be any
baryons in the present epoch. An alternative way of achieving a linear  
evolution of the scale factor is an effective
Einstein theory with a repulsive  
effective gravitational constant at long distances.
Such possibilities follow from effective gravitational actions
that have been considered in the past [10].
For a fourth order theory with action:
$$ 
S = \int d^4x\sqrt{-g}[\alpha R^2 - \beta R] \eqno{(14)}
$$
in the weak field approximation, the effective Newtonian potential
is:
$$\phi = - {a\over r} + b{exp(-\mu r)\over r} \eqno{(15)}$$
For $\mu r << 1$ we can have a canonical effective attractive theory. 
Over 
large distances, the effective potential is 
dominated by the first repulsive term
alone. A similar possibility occurs
in the conformally invariant higher order theory of gravity[11].
Choosing the gravitational action to be the square of the
Weyl tensor gives rise to an effective gravity action:

$$ 
S = \int d^4x\sqrt{-g}[\alpha C^2 - \beta R]\eqno{(16)}
$$
The dynamics of a conformally flat FRW 
metric is driven by the anomalous repulsive term alone. Canonical
attractive domains occur in the model as non - conformally flat 
perturbations in the FRW spacetime. 

     Yet another way of realising a linear evolution of the scale
factor is in a class of Brans - Dicke cosmological models [12].

    Linear evolution of the scale factor would also be possible
in the following ``toy'' model [13] that combines the Lee - Wick
construction of non - topological soliton [NTS] solutions [14] in
a variant of an effective gravity model proposed by Zee [15]. 
Consider the action:
$$
S = \int d^4x\sqrt{-g}[U(\phi)R + 
{1\over 2}\partial_\mu\phi\partial^\mu\phi - V(\phi) + L_m] \eqno{(17)} 
$$
Here $\phi$ is a scalar field non - minimally coupled to the
scalar curvature through the function$U(\phi)$, $V(\phi)$, its 
effective potential and $L_m$ the matter field action. $L_m$
includes a Higgs coupling of $\phi$ to a fermion. Let $V$ have a 
minimum at $\phi_{min}$ and a zero at $\phi^o$. We also choose
the Higg's coupling such that the effective fermion mass at
$\phi = \phi_{min}$ is greater than the effective fermion mass
at $\phi = \phi^o$. Finally we choose the non - minimal 
function $U(\phi_{min}) >> U(\phi^o)$. These conditions are
sufficient for the existence of large NTS's with the scalar
field trapped at $\phi = \phi^o$ in the interior of a large 
ball and quickly going to $\phi = \phi_{min}$ across the surface
of the ball. With a judicious choice of the surface tension, 
these balls could be as large as a typical halo
of a galaxy. The interior and exterior of such a ball 
would be regions with
effective gravitational constant 
$[U(\phi^o)]^{-1}~\&~[U(\phi_{min}]^{-1}$ 
respectively. With $[U(\phi_{min})]$ large enough, the universe 
would evolve as a curvature dominated dominated universe [without
any `K - matter']. 

     Such a universe would expand as a Milne universe having
canonical gravitating domains resticted to the interior of NTS 
domains. The interior would have a larger baryon entropy ratio, 
$\eta$, than the exterior. The requirement for the formation
and later spallation of $^4He$ defficient clouds onto a $^4He$
rich medium could be realised in such a model.

\vskip 2cm

Acknowledgment: Helpful discussions with Profs. Jim Peebles in ICGC, 
T. W. Kibble and G. W. Gibbons
are gratefully acknowledged.
 
\vfil\eject

\centerline{\bf References}

\item{1.} J. M. Escude \& M. J. Rees, Astro-ph 9701093 (1997) 
\item{2.} G. Steigman,  Astro-ph/9601126 (1996)
\item{3.} S. Weinberg, ``Gravitation and Cosmology'', John Wiley \&
sons, (1972).
\item{4.} See for example E. W. Kolb, and M. S. Turner, 
``The Early Universe'', Addison Wesley (1990).
\item{5.} L. Kawano, (1988), FERMILAB - PUB -88/34 and references therein.
\item{6.} R. V. Wagoner, Ap. J. Supp., $\underline{162}$, 18, 247 (1967)
\item{7.} R. I. Epstein, J. M. Lattimer \& D. N. Schram, Nature
$\underline{263}$, 198 (1976); R. Epstein, Astrophys. J. 
$\underline{212}$, 595 (1977); F. Hoyle \& W. Fowler, Nature
$\underline{241}$, 384 (1973)
\item{8.} E. W. Kolb, Astrophys. J., ${\underline 344}$, 543 (1989)
\item{9.} M. Sethi \& D. Lohiya,  ``Aspects of a coasting universe'',
GR15 proceedings (1997)
\item{10.} G. F. R. Ellis and M. Xu, 1995 (private communication).
\item{11.} P. Manheim and D. Kazanas, Gen. Rel \& Grav.
$\underline {22}$ 289 (1990)
\item{12.} H. Dehnen \& O. Obregon, Ast. and Sp. Sci. $\underline{17}$,
338 (1972); ibid. $\underline{14}$, 454 (1971)
\item{13.} D. Lohiya \& M. Sethi, ``A program for a problem free
cosmology within a framework of a rich class of scalar tensor
theories'', (1998)
\item{14.}(a) T. D. Lee, Phys. Rev. $\underline{D35}$, 3637 (1987);
T. D. Lee \& Y. Pang, phys. Rev. $\underline{D36}$, 3678 (1987);
(b) B. Holdom, Phys. Rev. $\underline{D36}$, 1000 (1987)
\item{15.} A. Zee, in ``Unity of forces in nature''
Vol II, ed. A. Zee, P 1082, World Scientific (1982);
Phys. Rev. Lett., $\underline{42}$, 417 (1979); 
Phys. Rev. Lett., $\underline{44}$, 703 (1980).

\vfill
\eject

\vskip 3cm

\centerline {\bf {TABLE I}}
\vskip 1cm
\centerline{
Abundances of Some Light Elements and Metals.} 

\vskip 2cm

\settabs 7 \columns
\+\bf$\eta$&$\bf^2 H$&$\bf^3 H$&
$\bf^3 He$&$\bf^4 He$&$\bf^7 Be$&$\bf^8 Li$ \& above \cr
\smallskip
\+$(10^{-9})$&$(10^{-18})$
&$(10^{-25})$&$(10^{-14})$&$(10^ {-1})$&$(10^{-11})$&$(10^{-8})$ \cr
\smallskip
\+$9.0$&$2.007$&$1.25$&$8.65$&$2.03$&$1.39$&$8.06$ \cr
\+$9.1$&$2.008$&$1.26$&$8.63$&$2.06$&$1.32$&$8.63$ \cr
\+$9.2$&$2.009$&$1.26$&$8.60$&$2.10$&$1.23$&$9.35$ \cr
\+$9.3$&$2.010$&$1.27$&$8.59$&$2.11$&$1.19$&$9.75$ \cr
\+$9.4$&$2.014$&$1.26$&$8.56$&$2.15$&$1.11$&$10.66$ \cr
\+$9.5$&$2.015$&$1.27$&$8.50$&$2.18$&$1.05$&$11.41$ \cr
\+$9.6$&$2.016$&$1.28$&$8.52$&$2.19$&$1.01$&$11.88$\cr
\+$9.7$&$2.017$&$1.28$&$8.49$&$2.22$&$0.96$&$12.69$ \cr
\+$9.8$&$2.020$&$1.29$&$8.47$&$2.25$&$0.91$&$13.51$ \cr
\+$9.9$&$2.020$&$1.29$&$8.45$&$2.28$&$0.86$&$14.47$ \cr
\+$10.0$&$2.020$&$1.30$&$8.43$&$2.30$&$0.83$&$15.19$ \cr

\bigskip
\noindent
\+Initial Temperature   $10^{11}K$ \cr
\+Final Temperature  $10^7K$ \cr 
\+No. of iterations at each step  550 \cr

\bye